\documentclass{natureprintstyle}

\usepackage{amsmath}
\usepackage{mathpazo} 
\usepackage{bm}
\usepackage{graphicx,subfigure,color}
\usepackage{fancyhdr}
\usepackage{pdfpages}
\usepackage{verbatim}
\usepackage[colorlinks,citecolor=blue]{hyperref} 
\usepackage[all]{hypcap}

\fancypagestyle{plain}{
\fancyhf{}
\fancyfoot[R]{\textit{Published in~}~N~A~T~U~R~E~~$|$~~9~~J~U~N~E~~2~0~1~6~~$|$~~ \thepage}

}
\def\hi{H{\sc i}}
\def\kms{km s$^{-1}$}

\newcommand\arcsec{\mbox{$^{\prime\prime}$}}%
\newcommand\farcs{\mbox{$.\!\!^{\prime\prime}$}}%

\def\lae{\mathrel{<\kern-1.0em\lower0.9ex\hbox{$\sim$}}}
\def\gae{\mathrel{>\kern-1.0em\lower0.9ex\hbox{$\sim$}}}

\def\mone{$^{-1}$}
\def\mtwo{$^{-2}$}

\newcommand{\Msol}{$M_{\odot}$}

\definecolor{Mygrey}{gray}{0.75}

\title{Cold, clumpy accretion onto an active supermassive black hole}
\author{Grant R.~Tremblay$^{1,2,\dagger}$, J.~B.~Raymond Oonk$^{3,4}$,
Fran\c{c}oise Combes$^{5}$, Philippe Salom\'{e}$^{5}$, Christopher P.~O'Dea$^{6,7}$, Stefi A.~Baum$^{6,8}$, G.~Mark Voit$^{9}$, Megan Donahue$^{9}$, 
Brian R.~McNamara$^{10}$, Timothy A.~Davis$^{11,2,\ddagger}$, 
Michael A.~McDonald$^{12}$, Alastair C.~Edge$^{13}$, ~~~~~~~Tracy E.~Clarke$^{14}$,
Roberto Galv\'{a}n-Madrid$^{15,2}$, Malcolm N.~Bremer$^{16}$,
Louise O.~V.~Edwards$^{1}$, Andrew C.~Fabian$^{17}$, Stephen Hamer$^5$, 
Yuan Li$^{18}$, Ana\"{e}lle Maury$^{19}$, Helen Russell$^{17}$, Alice C.~Quillen$^{20}$, C.~Megan Urry$^{1}$, Jeremy S.~Sanders$^{21}$, \& Michael Wise$^{3}$}

\begin{document}

\maketitle

\begin{abstract}  
Supermassive black holes in galaxy centres can grow by the
accretion of gas, liberating energy that might regulate star formation on
galaxy-wide scales\cite{1,2,3}. The nature of the gaseous fuel reservoirs that
power black hole growth is nevertheless largely unconstrained by observations,
and is instead routinely simplified as a smooth, spherical inflow of very hot
gas\cite{4}.  Recent theory\cite{5,6,7} and simulations\cite{8,9,10} instead
predict that accretion can be dominated by a stochastic, clumpy distribution
of  very cold molecular clouds --- a departure from the `hot mode' accretion
model --- although unambiguous observational support for this prediction
remains elusive. Here we report observations that reveal a cold, clumpy
accretion flow towards a supermassive black hole fuel  reservoir in the
nucleus of the Abell 2597 Brightest Cluster Galaxy (BCG),  a nearby (redshift
$\bm{z=0.0821}$) giant elliptical galaxy surrounded by a dense halo of hot
plasma\cite{11,12,13}. Under the right conditions, thermal instabilities can
precipitate from this hot gas, producing a rain of cold clouds that fall
toward the galaxy's centre\cite{14}, sustaining star formation amid a
kiloparsec-scale molecular nebula that inhabits its core\cite{15}.  The
observations show that these cold clouds also fuel black hole accretion,
revealing `shadows' cast by the molecular clouds as they move inward at about
$\bm{300}$ kilometres per second towards the active supermassive black hole in
the galaxy centre, which serves as a bright backlight. Corroborating evidence
from prior observations\cite{16} of warmer atomic gas at extremely high
spatial resolution\cite{17}, along with simple arguments based on geometry and
probability, indicate that these clouds are within the innermost hundred
parsecs of the black hole, and falling closer towards it.  
\end{abstract}

\begin{figure*}
\begin{center}
\includegraphics[width=0.8\textwidth]{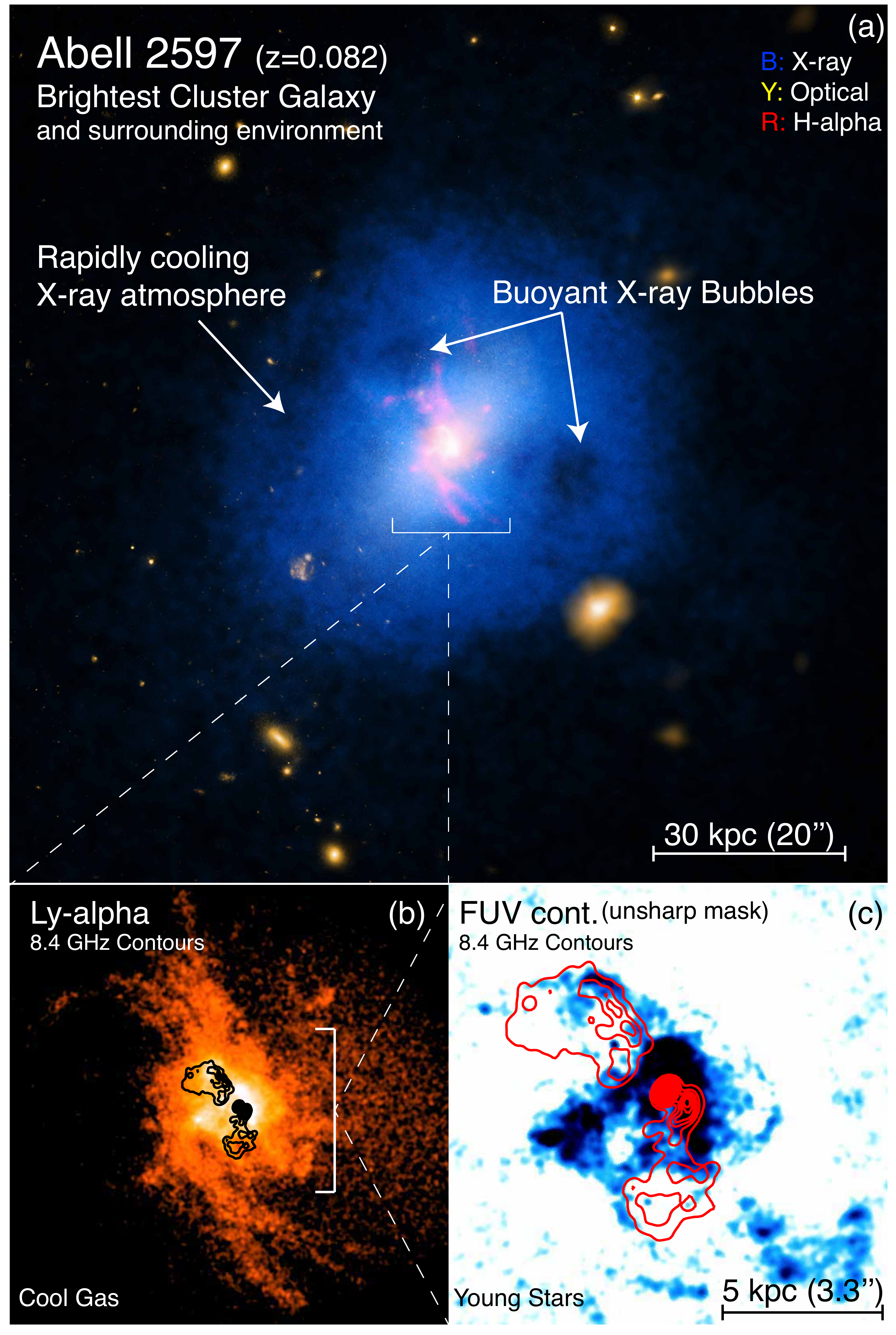}
\caption{{\textbf{A multiwavelength view of the Abell 2597 BCG.}} 
(a) {\it Chandra} X-ray, {\it HST} and DSS optical, and Magellan H$\alpha$+[N{\sc ii}] emission is shown in blue, yellow, and red, respectively (Credit: X-ray: NASA / CXC / Michigan State Univ / G.Voit et al; Optical: NASA/STScI \& DSS; H$\alpha$: Carnegie Obs. / Magellan / W.Baade Telescope / U.Maryland / M.McDonald). (b) {\it HST} image of Ly$\alpha$ emission associated with the ionised gas nebula13. (c) Unsharp mask of the HST far-ultraviolet continuum image of the central regions of the nebula\cite{13,18}. Very Large Array (VLA) radio contours of the 8.4 GHz source are overlaid in red.
\label{Fig1}}
\end{center}
\end{figure*}

\let\thefootnote\relax\footnote{
\begin{affiliations}
\item Yale Center for Astronomy \& Astrophysics, Yale University, 52 Hillhouse Ave., New Haven, CT 06511, USA
\item European Southern Observatory, Karl-Schwarzschild-Str. 2, 85748, Garching-bei-M\"unchen, Germany
\item ASTRON, Netherlands Institute for Radio Astronomy, P.O.~Box 2, 7990 AA Dwingeloo, The Netherlands
\item Leiden Observatory, Leiden University, Niels Borhweg 2, NL-2333 CA Leiden, The Netherlands
\item LERMA, Observatoire de Paris, PSL Research Univ., College de France, CNRS, Sorbonne Univ., Paris, France
\item Department of Physics \& Astronomy, University of Manitoba, Winnipeg, MB R3T 2N2, Canada
\item School of Physics \& Astronomy, Rochester Institute of Technology, 84 Lomb Memorial Dr., Rochester, NY 14623, USA
\item Chester F.~Carlson Center for Imaging Science, Rochester Institute of Technology, 84 Lomb Memorial Dr., Rochester, NY 14623, USA
\item Physics \& Astronomy Department, Michigan State University, East Lansing, MI 48824-2320, USA
\item Physics \& Astronomy Department, Waterloo University, 200 University Ave. W., Waterloo, ON, N2L, 2GL, Canada
\item School of Physics \& Astronomy, Cardiff University, The Parade, Cardiff CF24 3AA, United Kingdom
\item Kavli Institute for Astrophysics \& Space Research, MIT, 77 Massachusetts Ave., Cambridge, MA 02139, USA
\item Department of Physics, Durham University, Durham, DHL 3LE, United Kingdom
\item Naval Research Laboratory Remote Sensing Division, Code 7213 4555 Overlook Ave. SW, Washington, DC 20375, USA
\item Instituto de Radioastronom\'{i}a y Astrof\'{i}sica, UNAM, Apdo. Postal 3-72 (Xangari), 58089 Morelia, Michoac\'{a}n, M\'{e}xico
\item H.~W.~Wills Physics Laboratory, University of Bristol, Tyndall Avenue, Bristol, BS8 1TL, United Kingdom
\item Institute of Astronomy, Cambridge University, Madingly Rd., Cambridge, CB3 0HA, United Kingdom
\item Department of Astronomy, University of Michigan, 1085 S. University Avenue, Ann Arbor, MI 48109, USA
\item Laboratoire AIM-Paris-Saclay, CEA/DSM/Irfu CNRS --- Univ.~Paris Diderot, CE-Saclay, F-91191 Gif-sur-Yvette, France
\item Department of Physics \& Astronomy, University of Rochester, Rochester, NY 14627, USA
\item Max Planck Institut f\"{u}r Extraterrestrische Physik, 85748 Garching bei M\"{u}nchen, Germany \\ $\dagger$ Einstein Fellow $\ddagger$ Rutherford Fellow
\end{affiliations}
}

We observed the Abell 2597 Brightest Cluster Galaxy (Fig.~\ref{Fig1})  with
the Atacama Large Millimeter/submillimeter Array (ALMA), enabling  us to
create a three-dimensional map of both the location and motions  of cold gas
at uniquely high sensitivity and spatial resolution.  The ALMA receivers were
sensitive to emission from the $J=2-1$ rotational line of the carbon monoxide
(CO) molecule. CO(2-1) emission is used as a tracer of cold ($\sim 10-30$ K)
molecular hydrogen, which is vastly more abundant, but not directly observable
at these low temperatures.

\begin{figure*}
\begin{center}
\includegraphics[width=0.7\textwidth]{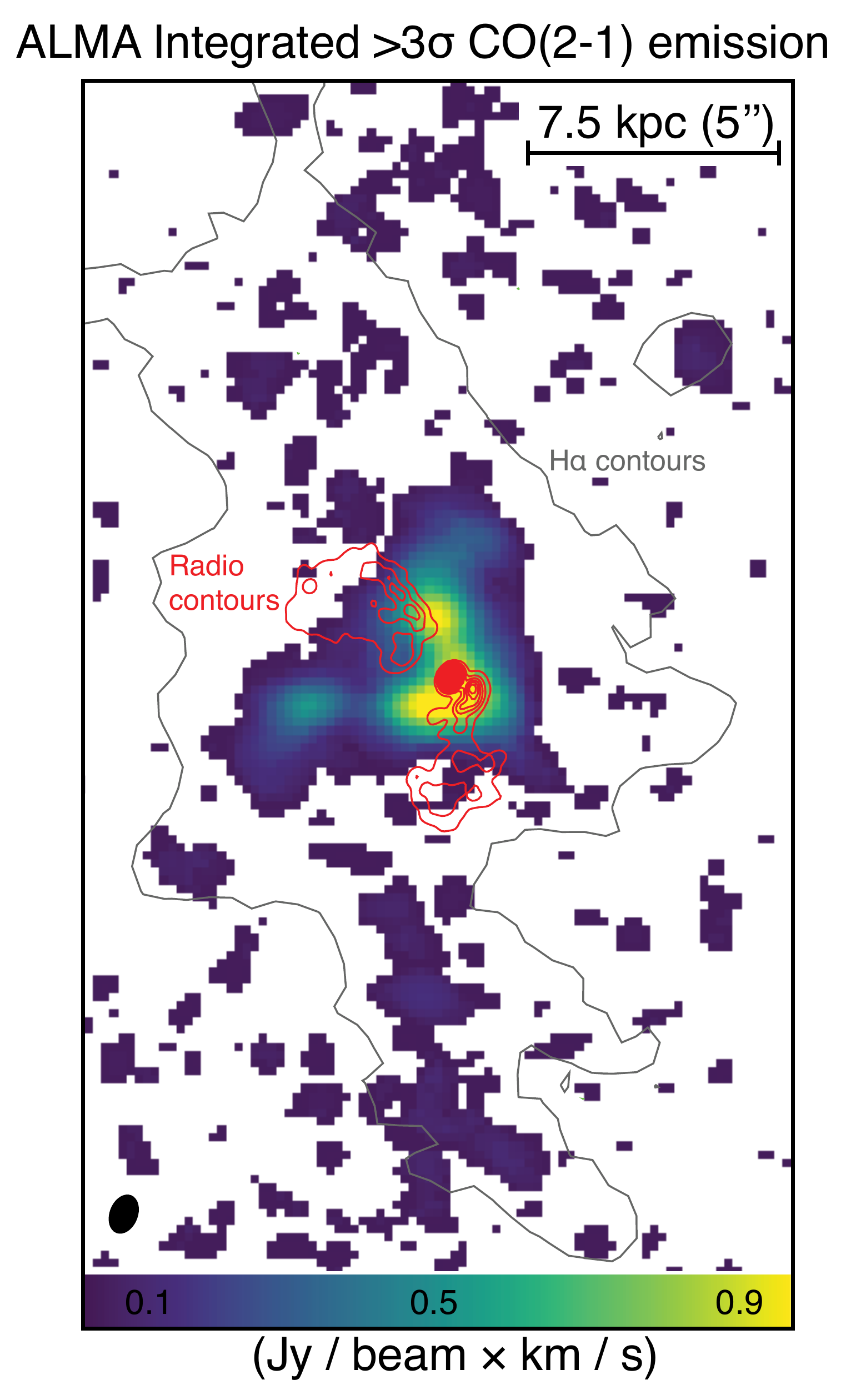}
\caption{{\textbf{ALMA observation of continuum-subtracted CO(2-1) emission in the Abell 2597 BCG.}} 
Emission is integrated from $-600$ to $+600$ \kms\ relative to the galaxy's systemic velocity. Channels are binned to $40$ \kms. Only $\ge3\sigma$ emission is shown. 8.4 GHz VLA radio contours are overlaid in black, and H$\alpha$+[N{\sc~ii}] contours outlining the rough boundary of the ionised nebula are shown in grey. The nebula is slightly larger than the grey contours suggest: emission outside of this boundary is still part of a smooth, fainter distribution of cold gas, cospatial with similarly faint emission in the optical.  
\label{Fig2}}
\end{center}
\end{figure*}

The continuum-subtracted CO(2-1) images (Fig.~\ref{Fig2}) reveal that  the
filamentary emission line nebula that spans the galaxy's innermost $\sim30$
kpc (Fig.~\ref{Fig1}b) consists not only of warm ionised gas\cite{18,19,20},
but cold molecular gas as well. In projection, the optical emission line
nebula is cospatial and morphologically matched with CO(2-1) emission detected
at a significance between $\gae 3\sigma$ (in the outer filaments) and
$\gae20\sigma$  (in the nuclear region) above the background noise level. The
warm ionised nebula is therefore likely to have a substantial molecular
component, consistent with results for other similar galaxies\cite{21}. The
total measured CO(2-1) line flux corresponds to a molecular hydrogen gas mass
of $M_{\mathrm{H}_2} = \left(1.8\pm0.2\right) \times 10^9$ \Msol, where \Msol\
is the mass of the sun. The critical (minimum) density for CO(2-1) emission
requires that the volume filling factor of this gas be very low, of order a
few percent. The projected spatial coincidence of both the warm ionised and
cold molecular nebulae therefore supports the long-envisaged hypothesis that
the ionised gas is merely the warm `skin' surrounding far colder and more
massive molecular cores\cite{22,23}, whose outer regions are heated by intense
radiation from the environment in which they reside. Rather than a monolithic,
kiloparsec-scale slab of cold gas, we are more likely observing a projected
superposition of many smaller, isolated clouds and filaments.

\begin{figure*}
\begin{center}
\includegraphics[width=0.8\textwidth]{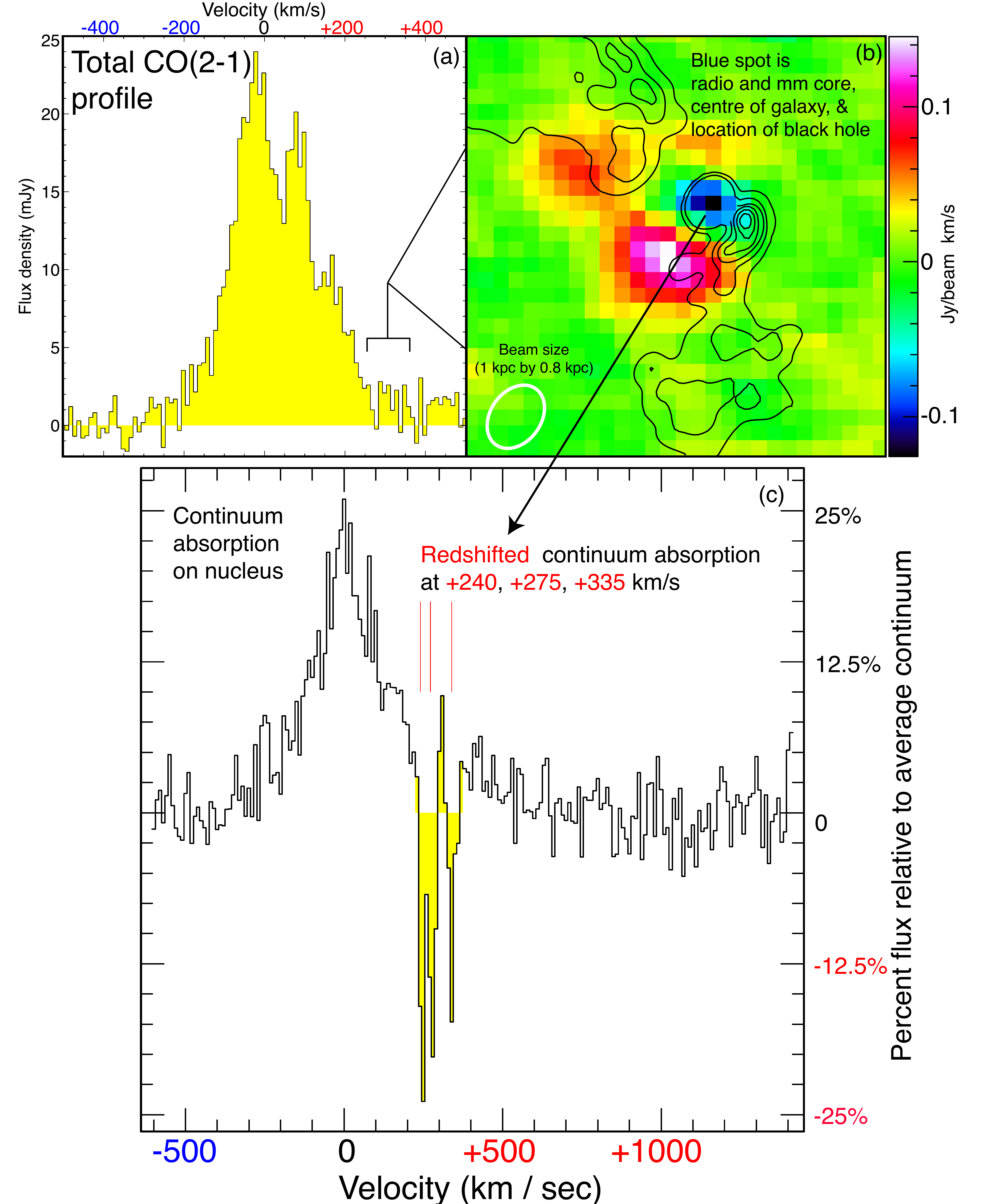}
\caption{{\textbf{`Shadows' cast by molecular clouds moving toward the supermassive black hole.}} (a) Continuum-subtracted ALMA CO(2-1) spectrum extracted from a central 10 kpc region. Brackets mark CO(2-1) emission shown in panel (b), where 8.4 GHz radio contours are overlaid. The central radio contours have been removed to aid viewing of the continuum absorption, seen as the blue/black spot of `negative' emission. (c) Continuum-subtracted CO(2-1) spectrum extracted from this region cospatial with the mm and radio core. Absorption lines are indicated in red. 
\label{Fig3}}
\end{center}
\end{figure*}

The data unambiguously show that cold molecular gas is falling inward  along a
line of sight that intersects the galaxy centre. We know this  because the
ALMA beam cospatial with the millimetre continuum source,  the radio core, and
the isophotal centre of the galaxy reveals strong,  redshifted continuum
absorption (Fig.~\ref{Fig3}b), found by extracting the CO(2-1) spectrum from
this central beam. This reveals at least three deep and narrow absorption
lines (Fig.~\ref{Fig3}c), with redshifted line centres at $+240$, $+275$, and
$+335$ \kms\ relative to the systemic (stellar) velocity of the galaxy, all
within an angular (physical) region of  $0\farcs715 \times 0\farcs533$
($1~\mathrm{kpc}\times0.8~\mathrm{kpc}$).

\begin{figure*}
\begin{center}
\includegraphics[width=0.8\textwidth]{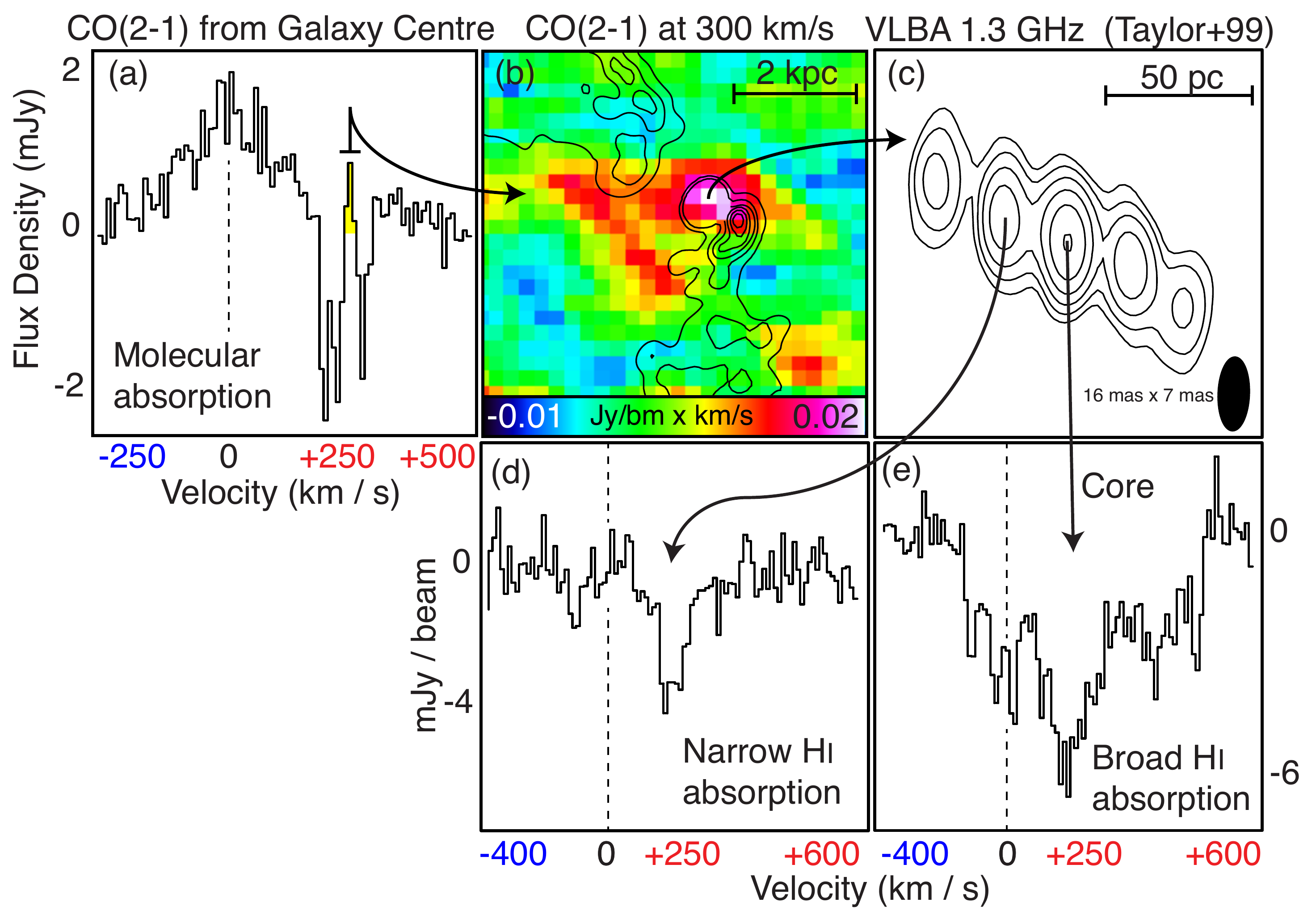}
\vspace*{-3mm}
\caption{{\textbf{Corroborating evidence that the inflowing molecular clouds must be in close proximity to the black hole.}} (a)  CO(2-1) absorption spectrum from Fig.~\ref{Fig3}, with a region of emission at $\sim +300$ \kms\ marked in yellow. (b) Integrated CO(2-1) emission from this region (colour coded), showing that gas at $\sim +300$ \kms\ is confined to the innermost $\sim2$ kpc of the galaxy. (c) 1.4 GHz radio continuum source from an archival VLBA observation\cite{17} with an extremely high physical resolution of $\sim25$ parsecs by $\sim10$ parsecs. (d,e) \hi\ 21 cm absorption observed against this synchrotron jet. The signal varies dramatically over tens of parsec scales.   
\label{Fig4}}
\end{center}
\end{figure*}

These absorption features arise from cold molecular clouds moving  toward the
centre of the galaxy, either via radial or inspiralling trajectories.  They
manifest as continuum absorption because they cast `shadows' along the line of
sight as the clouds eclipse or attenuate about $~20\%$ (or about 2 mJy)  of
the millimetre synchrotron continuum source, which serves as a bright
backlight (13.6 mJy at rest-frame 230 GHz). The synchrotron continuum is
emitted by jets launched from the accreting supermassive ($\sim3\times10^{8}$
\Msol)\cite{13} black hole in the galaxy's active nucleus (Fig.~\ref{Fig4}).
The absorbers must therefore be located somewhere between the observer and the
galaxy centre, falling deeper into the galaxy at $\sim +300$ \kms\  toward the
black hole at its core.  This radial speed is roughly equal to the expected
circular velocity\cite{24} in the nucleus, consistent either with a nearly
radial orbit,  or highly non-circular motions in close proximity to the
galaxy's core.

Gaussian fits to the spectral absorption features reveal narrow linewidths of
$\sigma_v \lae 6$ \kms,  which means the absorbers are more likely spatially
compact,  with sizes that span tens (rather than hundreds or thousands) of
parsecs.  The shapes of the absorption lines remain roughly the same
regardless of how finely the spectra are binned, suggesting that the absorbers
are likely coherent structures, rather than a superposition of many smaller
absorbers unresolved in velocity space. If each absorption feature corresponds
to one coherent cloud, and if those clouds roughly obey size-linewidth
relations\cite{25,26} for giant molecular clouds in the Milky Way, they should
have diameters not larger than $\sim40$ pc. If in virial equilibrium,
molecular clouds this size would have masses of order $10^{5-6}$ \Msol, and if
in rough pressure equilibrium with their ambient multiphase $10^{3-7}$ K
environment\cite{13}, they must have high column densities of order
$N_{\mathrm{H}_2}\approx 10^{22-24}$ cm\mtwo\  so as to maintain pressure
support. The thermal pressure in the core of Abell 2597 is nearly three
thousand times\cite{11} greater than that for the Milky Way,  however, which
means the absorbing clouds may be much smaller.

The absorbers have optical depths that range from
$0.1 \lae \tau_{\mathrm{CO(2-1)}} \lae 0.3$.
The physical resolution of the ALMA data is larger than the synchrotron
background source, which means that the optical depth is likely contaminated
by an unresolved, additive superposition of both emission and absorption
within the beam. Compact, dense cold clouds are nevertheless likely to be
optically thick, which may mean they eclipse the continuum source with an
optical depth of unity but a small covering factor of roughly 0.2. Especially
when considering beam contamination by emission, the covering factor cannot be
known with certainty, as this depends on the unknown geometry of the absorbing
and emitting regions within the ALMA beam.

This geometry can be constrained, however, given existing Very Long Baseline
Array (VLBA) radio observations at extremely high spatial resolution\cite{15}.
These data resolve the 1.3 and 5 GHz radio continuum source down to scales of
25 parsecs, revealing a highly symmetric, 100 pc-scale jet about a bright
radio core (Fig.~\ref{Fig4}c). Just as we have found in cold molecular gas,
inflowing warmer atomic hydrogen gas (\hi) has previously been found in
absorption against this pc-scale jet, corroborating prior reports of inflowing
atomic gas at lower spatial resolutions\cite{14}. The inflow velocity of this
gas matches that seen in our ALMA data. Remarkably, both the optical depth
and linewidth of the warm atomic absorption signal varies dramatically  across
the jet, with a broad ($\sigma_v\approx310$ \kms) component cospatial with the
core that is absent just $\sim20$ pc to the northeast, where only a narrow
($\sigma_v\approx50$ \kms) \hi\ line is found at the same redshift. This
effectively requires the inflowing atomic gas to be confined within the
innermost $\sim100$ pc of the black hole, as gas further out would give rise
to an unchanging absorption  signal across the compact jet. The infall
velocity is the same  as that for the cold molecular clouds seen in CO(2-1)
absorption, which means they most likely stem from the same spatial region,
within tens of parsecs of the accreting black hole.

This is further supported by the ALMA data itself.  In emission, all gas
around $\sim+300$ \kms\ that is conceivably available to attenuate the
continuum signal is confined to the  innermost 2 kpc about the nucleus
(Fig.~\ref{Fig4}a,b). The radial dependence of molecular cloud volume number
density within this region is  uncertain, but probably steeper than $r^{-1}$,
and likely closer to $r^{-2}$ (Fig.~\ref{Fig4}b). This means that the chances
of a random line  of sight crossing will drop with increasing distance from
the black hole.  If the gas volume density goes as $r^{-2}$, a cloud 100 pc
from the black hole is ten times more likely to cross our line of sight than a
cloud at a galactocentric distance of 1 kpc. It would be exceedingly unlikely
for three such clouds to cross our line of sight to the black  hole were they
spread over several kiloparsecs throughout the galaxy's outskirts.

The data therefore serve as strong observational evidence for an  inward-
moving, clumpy distribution of molecular clouds within a few hundred parsecs
of an accreting supermassive black hole.  The infalling clouds are likely a
few to tens of parsecs across and therefore massive (perhaps $10^{5-6}$ \Msol\
each). If they are falling directly toward the black hole, rather than bound
in a non-circular orbit that tightly winds around it, they could supply an
upper-limit accretion rate on the order of $\sim0.1$ to a few \Msol\ yr\mone,
depending on the three dimensional distribution of infalling clouds. If most
of the clouds are instead locked in non-circular orbits around the black hole,
the fuelling rate would depend on   the gas angular momentum, and the local
supply of torques that might lessen it. Simulations suggest\cite{9,10,14} that
such torques may be plentiful, as they predict a stochastic `rain' of thermal
instabilities that condense from all directions around the black hole,
promoting angular momentum cancellation via tidal stress and cloud-cloud
collisions. Even highly elliptical cloud orbits should therefore be associated
with significant inward radial motions. The clouds might fall onto the
accretion disc itself, or into a clumpy rotating ring akin to the `torus'
invoked in AGN unification models\cite{27}.

Cold accretion onto black holes has long been predicted by both theory and
simulations\cite{5,6,7,8,9,10,14}, but it has not been definitively observed in a
manner so stripped of ambiguity regarding the clouds' proximity to a black
hole. While no observation of a single galaxy can prove this theoretical
prediction to be definitively true, the combined ALMA and VLBA dataset for
Abell 2597 enable a uniquely unambiguous observation of molecular clouds that
are either directly associated with black hole growth, or are soon about to
be. The result augments a small but growing set of previously published
molecular absorption systems\cite{28,29,30} in which black hole proximity is
less well constrained. These could nevertheless be used to inform future
systematic searches for cold black hole accretion across larger samples.
Multi-epoch observations with ALMA might reveal shifts in the absorption
lines, confirming their close proximity, and resolving cold black hole
accretion as it evolves through time.

\noindent
\textbf{\textsf{\footnotesize Received 17 December 2015; accepted 22 March 2016.}}\\
\textbf{\textsf{\footnotesize Published in \textit{Nature}, Volume 534, Issue 7607, Pages 218-221.}}\\
\textbf{\textsf{\footnotesize DOI: \href{http://www.nature.com/articles/doi:10.1038/nature17969}{10.1038/nature17969}}}\\

\begin{addendum} 

\item  This paper makes use of the following ALMA data:
ADS/JAO.ALMA\#2012.1.00988.S. ALMA is a partnership of ESO (representing its
member states), NSF (USA) and NINS (Japan), together with NRC (Canada) and NSC
and ASIAA (Taiwan), in cooperation with the Republic of Chile. The Joint ALMA
Observatory is operated by ESO, AUI/NRAO and NAOJ. We are grateful to the
European ALMA Regional Centres, particularly those in Garching and Manchester,
for their dedicated end-to-end support of data associated with this paper. We
thank Prof.~Richard Larson for discussions. G.R.T.~acknowledges support from
the National Aeronautics and Space Administration (NASA) through Einstein
Postdoctoral Fellowship Award Number PF-150128, issued by the Chandra X-ray
Observatory Center, which is operated by the Smithsonian Astrophysical
Observatory for and on behalf of NASA under contract NAS8-03060. F.C.
acknowledges the European Research Council (ERC) for the Advanced Grant
Program \#267399-{\emph{Momentum}}. B.R.M.~is supported by a generous grant
from the Natural Sciences and Engineering Research Council of Canada.
T.A.D.~acknowledges support from a Science and Technology Facilities Council
(STFC) Ernest Rutherford Fellowship. A.C.E.~acknowledges support from STFC
grant ST/L00075X/1. A.C.F. and H.R.R. acknowledge support from ERC Advanced
Grant Program \#340442-{\emph{Feedback}}. M.N.B.~acknowledges funding from the
STFC. Basic research in radio astronomy at the Naval Research Laboratory is
supported by 6.1 Base funding.

 \item[Author Contributions] G.R.T.~was principal investigator on the original proposal, 
 performed the data analysis, and wrote the paper. J.B.R.O., T.A.D., R.G.M., and A.M.~were 
 substantially involved in planning both scientific and technical aspects of the proposal, 
 while T.A.D.~and R.G.M.~contributed ALMA data reduction and analysis expertise once the data were obtained. 
 J.B.R.O., F.C., and P.S.~invested substantial time in analysis of the data. Substantial scientific 
 feedback was also provided over many months by F.C., J.B.R.O., C.P.O., S.A.B., G.M.V., M.D., B.R.M., M.A.M., 
 T.E.C., H.R., A.C.E. and A.C.F., while all other co-authors discussed the results and commented on the manuscript.

 \item[Author Information] Reprints and permissions information is available at
www.nature.com/reprints. The authors declare no competing financial interests.
Readers are welcome to comment on the online version of the paper.
Correspondence and requests for materials
should be addressed to G.R.T.~(grant.tremblay@yale.edu).

\end{addendum}

\clearpage

\begin{methods}

\subsection{Observations, Data Reduction, and Analysis.}  

The new ALMA data presented in this paper were  obtained in Cycle 1 with the
use of 29 operational antennae in the 12m Array.  ALMA's Band 6 heterodyne
receivers were tuned to a frequency of 213 GHz, sensitive  to the $J=2-1$
rotational line transition of carbon monoxide at the redshift of the Abell
2597 BCG ($z=0.0821$). The ALMA correlator, set to Frequency Division Mode
(FDM), delivered a bandwidth of 1875 MHz (per baseband) with a 0.488 MHz
channel spacing, for a maximum spectral resolution of $\sim2$ km s\mone. One
baseband was centered on the CO(2-1) line, while the  other three sampled the
local continuum.  Maximum antenna baselines extended to $\sim1$ km, delivering
an angular resolution at 213 GHz of $\sim 0\farcs7$ within a $\sim 28\arcsec$
primary beam (field of view). ALMA observed the Abell 2597 BCG for a total of
$\sim 3$ hours over three separate scheduling blocks executed between 17-19
November 2013. The planet Neptune and quasars J2258-2758 and J2331-1556 were
used for amplitude, flux, and phase calibration.  The data were reduced using
{\sc casa} version 4.2 with calibration and imaging scripts kindly provided by
the ALMA Regional Centres (ARCs) in both Garching, Germany, and Manchester,
UK. Beyond the standard application of the phase calibrator solution, we
iteratively performed self-calibration of the data using the galaxy's own
continuum,  yielding a $\sim 14\%$ decrease in RMS noise to a final value of
$0.16$ mJy  per $0\farcs715\times0\farcs533$  beam per 40 km~s\mone\ channel.
There is effectively no difference in CO(2-1) morphology  between the self-
calibrated  and non-self-calibrated cubes. Measurement sets were imaged using
`natural' visibility weighting and binning to either 5 km s\mone, 10 km s\mone
or 40 km s\mone, as indicated  in the figure captions. The figures presented
in this paper show only continuum-subtracted, pure CO(2-1) line emission. The
rest-frame 230 GHz continuum observation is dominated by a bright (13.6 mJy)
point source associated with the AGN (detected at $\gae400\sigma$), serving as
the bright `backlight' against which the continuum absorption features
presented in this Letter were observed. The continuum data also features
compact ($\sim 5$ kpc) extended emission at $\sim 10\sigma$ that extends along
the galaxy's dust lane, to be discussed in a forthcoming paper.  \\

\subsection{Adoption of a systemic velocity.}

Interpretation of gas motions relative to the stellar component of a galaxy
requires adoption of a systemic (stellar) velocity to be used as a `zero
point' marking the transition from blue- to redshift. All CO(2-1) line
velocities discussed in this Letter are set relative to 213.04685 GHz, where
observed CO(2-1) emission peaks. This frequency corresponds to $^{12}$CO(2-1)
(rest-frame 230.538001 GHz) at a redshift of $z=0.0821$. This redshift is
consistent, conservatively within $\pm60$ \kms, with every other available
multiwavelength tracer  of the galaxy's systemic velocity, including prominent
Ca{\sc~ii} {\sc h}, {\sc k}, and G-band absorption features\cite{16} that
directly trace the galaxy stellar component, the redshift of all optical
emission lines\cite{31}, as well as a broad (FWHM $\sim412$ \kms) \hi\
absorption component\cite{16} at the optical emission and absorption line
redshift. It is also consistent, within $\pm60$ \kms, with a cross-correlation
of emission and absorption lines using galaxy template spectra\cite{15}, as
well as with all other published reports of the galaxy's systemic velocity
(found, e.g., within the HyperLeda database). We are therefore certain that
the reported redshift of the absorption features discussed in this letter
indeed corresponds to real motion relative to the galaxy's stellar component.
Without caveat or ambiguity, the absorbing cold clouds are moving {\it into}
the galaxy at roughly $\sim +300$ \kms. \\

\subsection{Mass Estimates.} 
All molecular gas masses estimated in this letter adopt the following relation\cite{32}:

\begin{equation}
\begin{split}
M_\mathrm{mol}  = & \frac{1.05 \times 10^4}{3.2} ~ \left(\frac{X_{\mathrm{CO}}}{2\times 10^{20}~\frac{\mathrm{cm}^{-2}}{\mathrm{K~km~s}^{-1}}}\right) \\ & \times \left( \frac{1}{1+z}\right) \left(\frac{S_{\mathrm{CO}}\Delta v}{\mathrm{Jy~km~s}^{-1}}\right) \left(\frac{D_\mathrm{L}}{\mathrm{Mpc}}\right)^2 M_\odot, 
\end{split}
\end{equation}

where $S_{\mathrm{CO}}\Delta v$ is the emission integral for CO(1-0) (effectively the total CO(1-0) flux over the region of interest), 
$z$ is the galaxy redshift ($z=0.0821$), and $D_L$ its luminosity distance (373.3 Mpc), for which 
we assume a flat
$\Lambda$CDM model wherein  $H_0 =  70$ km  s$^{-1}$ Mpc$^{-1}$,
$\Omega_M =  0.3$, and $\Omega_{\Lambda} =  0.7$.
This mass estimate most critically relies on an assumption of the CO-to-H$_2$ conversion factor\cite{32} $X_{\mathrm{CO}}$.
In this Letter we assume the average Milky Way value of $X_{\mathrm{CO}} = 2 \times 10^{20}$ cm\mtwo\ $\left(\mathrm{K~km~s}^{-1}\right)^{-1}$
and a CO(2-1) to CO(1-0) flux density ratio of $3.2$. Other authors have provided extensive discussion of these assumptions as they 
pertain to cool core BCGs\cite{30,33,34}. Scientific conclusions 
in this paper are largely insensitive to choice of $X_{\mathrm{CO}}$. \\

A single gaussian fit to the CO(2-1) spectrum extracted from an aperture 
containing all detected emission yields an 
emission integral of $S_{\mathrm{CO}}\Delta v = 4.2\pm0.4$ Jy km s\mone\ with a line FWHM of $252 \pm 14$ km s\mone,  
corresponding to a total molecular hydrogen (H$_2$)
gas mass of $M_{\mathrm{H}_2}  = \left(1.80 \pm 0.19\right) \times 10^9$ \Msol.
This is very close to the previously reported\cite{13} mass, based on an IRAM 30m CO(2-1) observation, 
of $\left(1.8\pm0.3\right) \times 10^9$ \Msol. This comparison is not one-to-one, as 
the mass from the IRAM 30m observation was computed within a beam size of 
11\arcsec\ (rather than 28\arcsec for the ALMA data), and used a CO(2-1)/CO(1-0) flux ratio of 4 (rather than 3.2, as we use here). 
These differences are minor, particularly because nearly all of the CO(2-1) emission detected by ALMA is found 
within the central 11\arcsec\ size of the IRAM 30m beam. 
It is therefore safe to say that our ALMA observation has 
detected nearly all emission that was detected in the single-dish IRAM 30m observation, and that 
very little extended emission has been `resolved out'. \\

\subsection{Estimating physical properties of the redshifted absorbing gas.}

We have estimated a rough upper-limit size of the absorbing clouds
assuming the widely-adopted Larson et al.\cite{25} and Solomon et al.\cite{26}~size-linewidth relation for molecular clouds in the Milky Way (namely, the Solomon et al.~1987 fit),
\begin{equation}
\sigma_v = \left(1.0 \pm 0.1 \right) S^{0.5\pm0.05}~\mathrm{km~s}^{-1}, 
\end{equation}
where $\sigma_v$ is the velocity line-width of the cloud 
and $S$ is the diameter of the cloud in parsecs. \\

A measured absorber linewidth of  
$\sigma_v \sim 6$ km s\mone\ would then correspond to a size of $\sim 36$ pc. 
As noted in the main text of the Letter, the thermal pressure  in the Abell
2597 Brightest Cluster Galaxy is roughly three thousand times
higher than that for the Milky Way,   so it is likely that the above relation
does not apply. A higher   ambient pressure implies higher compression and
therefore smaller cloud size,   so the above estimate should, at best, be
considered a very rough upper-limit. The main lesson to take away 
from this exercise is that the absorbing clouds are likely physically compact
(i.e. a few to tens of parsecs in diameter, rather than hundreds of pc). \\

The three clouds are separated from one another by $\sim45-60$ km s\mone\ in 
velocity space, which means they are unlikely to be closely bound satellites of 
one another. Instead, it is more likely that they represent three random 
points along a radial distribution of clouds. \\

If the absorbers are in virial equilibrium, their masses $M_{\mathrm{cloud}}$ can be roughly 
estimated by applying the virial relation, 
\begin{equation}
\begin{split}
M_{\mathrm{cloud}} & \approx \frac{R_{\mathrm{cloud}} \sigma_v^2}{G} \\ & \approx \frac{20~\mathrm{pc} \times \left( 6~\mathrm{km/s} \right)^2}{4.302 \times 10^{-3}~\mathrm{pc}~M_\odot^{-1} \left( \mathrm{km/s} \right)^2} \\ & \approx 1.7\times10^5~M_\odot,
\end{split}
\end{equation}
where $R_{\mathrm{cloud}}$ is the cloud radius (as roughly estimated above)
and $\sigma_v$ is its velocity dispersion (also as above). \\

CO(2-1) optical depths for the absorbers were estimated 
by assuming that 
\begin{equation}
I_{\mathrm{total}} = I_{\mathrm{continuum}}\mathrm{e}^{-\tau_{\mathrm{CO(2-1)}}},
\end{equation}
where $I_{\mathrm{total}}$ and $I_{\mathrm{continuum}}$ are the integrated 
intensities of the total (line plus continuum) and continuum-only signals, respectively, and $\tau_{\mathrm{CO(2-1)}}$ is the optical depth of the CO(2-1) absorption feature. \\

The stellar velocity dispersion of the BCG\cite{24} is $\sigma_v = 220 \pm
19$ km s\mone.  Under the assumption of an isothermal sphere, the circular
velocity should be $\sim 300$ km s\mone, (i.e., $\sqrt{2}\sigma_v$) which
is roughly  the line of sight velocity of the absorption features.
The redshift of the absorption features is a significant fraction of this, 
which means they could be on a purely radial orbit (though their transverse 
velocity cannot be known with this single observation). \\

If our line of sight is representative, and therefore a `pencil beam' sample 
of a three-dimensional spherical distribution of clouds, 
the total mass of cold gas contained within this distribution should go roughly as  
\begin{equation}
M \approx 10^9 M_\odot \times f_c \times \left( \frac{r}{1~\mathrm{kpc}}\right)^2 \times \left(\frac{N_\mathrm{H}}{10^{22}~\mathrm{cm}^{-2}}\right)
\end{equation}
where $f_c$ is the covering factor and $r$ is the radius of an imaginary 
thin spherical shell of molecular gas with column density $N_{\mathrm H}$. If such a shell had a covering factor of 1, a radius of 1 kpc, and 
a column density of $10^{22}$ cm\mtwo, then the total mass of molecular hydrogen 
contained within that shell would be roughly one billion solar masses. 
A column density in excess of $10^{22}$ requires this distribution to be 
contained within a sphere of radius $<<1$ kpc, lest the total mass
of molecular hydrogen in the galaxy be violated. 
If the characteristic column density is $10^{23}$ cm\mtwo, for example, 
this mass must be contained within a sphere of radius 300 pc, or else its total mass would exceed the $\sim 1.8\times10^9$ \Msol\ present in the system. \\

\subsection{Code, software, and data availability.}

The raw ALMA data used in this Letter are publicly available at
the ALMA Science Archive, accessible here: \url{https://almascience.nrao.edu/aq/} (search for project code 2012.1.00988.S). 
Codes that we have written to both reduce and analyse the ALMA data have been made publicly available here: \url{https://github.com/granttremblay/Tremblay_Nature_ALMA_Abell2597}. Reduction of the data as well as some simple modeling (e.g., fitting of Gaussians to lines) was performed using routines included in CASA version 4.2, available here: \url{http://casa.nrao.edu/casa_obtaining.shtml}. This research made use of Astropy (\url{http://www.astropy.org}), a community-developed core Python package for Astronomy\cite{35}.

\end{methods}

\end{document}